%% file: z_final.tex
\documentclass[aps,prl,twocolumn,superscriptaddress,showpacs]{revtex4} % UTOBS
\newcommand{\figsize}{0.36\textwidth}      %              UTOBS
\usepackage{graphicx}% Include figure files
\usepackage{dcolumn}% Align table columns on decimal point
\usepackage{bm}% bold math
\usepackage{setspace}
\def\met{\mbox{${\hbox{$E$\kern-0.6em\lower-.1ex\hbox{/}}}_T$}} %missing ET
\def\mex{\mbox{${\hbox{$E$\kern-0.6em\lower-.1ex\hbox{/}}}_x$}} %missing Ex
\def\mey{\mbox{${\hbox{$E$\kern-0.6em\lower-.1ex\hbox{/}}}_y$}} %missing Ey
\def\mexy{\mbox{${\hbox{$E$\kern-0.6em\lower-.1ex\hbox{/}}}_{x,y}$}} %missing Exy
 
\def\njet{N_{\rm jet}}                   %N_jet
\def\njetcor{N_{\rm jet}^{\rm cor}}  
\def\sigmanjet{\sigma_{N_{\rm jet}}}
\def\D0{D\O}                            %D0
\newcommand{\ptjet}{ p_{\rm T}^{\rm jet}}

\newcommand{\ptjetcor}{p_{\rm T, cor}^{\rm jet}}

\newcommand{\yjet}{y^{\rm jet}}
\newcommand{\ptjetcal}{p_{\rm T,cal}^{\rm jet}}

\newcommand{\yjetcal}{y^{\rm jet}_{\rm cal}}
\newcommand{\fjetcal}{\phi^{\rm jet}_{\rm cal}}
\newcommand{\deltapt}{  \delta_{p_{\rm  T}}^{\rm  mi}}

\newcommand{\fake}{f^{\rm jet}_{\rm e}}

\def\Zee{\mbox{$Z/\gamma^* (\rightarrow e^+ e^-)$}}
\def\Ztt{\mbox{$Z/\gamma^* (\rightarrow \tau^+ \tau^-)$}}

\def\tt{\mbox{$t \overline{t}$}}
\begin{document}
%
% ======> Title of the paper goes here <====================
%
\title{
Measurement of Inclusive Jet Cross Sections in $\Zee$+jets Production \\ in 
{\boldmath $p\overline{p}$} Collisions at 
{\boldmath $\sqrt{s}$} = 1.96 TeV \\
%(version v1 \ \today)
}
\input{July_2007_Authors_Visitors}
% --------- End of author list and affiliations ----------

\date{\today}

%
% ==============> Text of the abstract goes here <=====================
% 
\begin{abstract}

Inclusive jet cross sections in $Z/\gamma^*$ events, with $Z/\gamma^*$ decaying into an electron-positron pair, are measured as a 
function of jet transverse momentum and jet multiplicity 
in $p \overline{p}$ collisions at 
$\sqrt{s} = 1.96 \ {\rm TeV}$ with the upgraded Collider Detector 
at Fermilab in Run II, based on an integrated luminosity of $1.7 \ \rm fb^{-1}$.
The measurements cover the rapidity region
$| \yjet | < 2.1$
and the transverse momentum range \mbox{$\ptjet > 30$~GeV/c}.
Next-to-leading order perturbative QCD predictions are in good agreement with the 
measured cross sections.

\end{abstract}

\maketitle

%\vfill\eject
%\twocolumn

% ======================
%\section{Introduction}
% ======================

The measurement of the inclusive production of collimated jets of hadrons 
in association with a $Z/\gamma^*$ boson in 
$p \overline{p}$ collisions provides a stringent 
test of perturbative quantum chromodynamics (pQCD)~\cite{pqcd}. 
The understanding of $Z/\gamma^*$+jets final states is a crucial 
part of the physics program at the Tevatron since     
they constitute important irreducible backgrounds in searches for 
new physics. 
Previous  results~\cite{cdfrunI} from Run~I at
the Tevatron have been compared with leading-order (LO) 
plus parton shower Monte Carlo predictions affected by large scale uncertainties.
This Letter reports new and more precise measurements of the inclusive jet cross sections in
$\Zee$  production  
using 1.7 $\rm fb^{-1}$ of data collected by the CDF
experiment in Run~II. Inclusive jet differential cross sections as a function of jet transverse momentum $\ptjet$~\cite{coord}
and total cross sections as a function of jet multiplicity $\njet$ are measured.  
The data are  compared to next-to-leading 
order (NLO) pQCD predictions~\cite{mcfm} 
 including non-perturbative contributions. 

% =================
%\section{CDF Detector}  (DONE)
% ======================

The CDF II detector is described in detail elsewhere~\cite{cdfii}.
The detector has a charged particle tracking system immersed in a 1.4~T magnetic field 
aligned coaxially with the beam line that provides tracking coverage in the 
pseudorapidity~\cite{coord} range $|\eta| \leq 2$. 
%
% A silicon microstrip detector provides tracking 
% over the radial range 1.35 to 28 cm  and covers the pseudorapidity~\cite{coord} range 
% $|\eta| \leq 2$.  A cylindrical 3.1 m long open-cell drift chamber covers the radial range 
% from 44 to 132 cm and provides full tracking coverage for $|\eta| \leq 1$. 
%
Segmented sampling calorimeters, arranged in a projective tower
geometry, surround the tracking system and measure the energy of 
interacting particles for $|\eta| < 3.6$. The central electromagnetic and hadronic calorimeters~\cite{ccal}
cover the region $|\eta| < 1$,  
% It consists of two sections, an electromagnetic calorimeter 
% and a hadronic calorimeter, divided into 480 towers of size $0.1$ in $\eta$ and  
% $0.26$ in $\phi$. 
while the end-wall hadronic calorimeter~\cite{wha} 
%is behind the central 
%barrel calorimeter in the region $0.6 < |\eta| < 1.0$, 
provides forward coverage out to 
$|\eta| < 1.3$. Forward  electromagnetic and hadronic 
calorimeters~\cite{pcal} cover the regions $1.1 < |\eta| < 3.6$ and $1.3 < |\eta| < 3.6$, respectively.
%
% In Run II, new forward scintillator-plate calorimeters~\cite{pcal} replaced 
% the Run I gas calorimeter system. The new plug electromagnetic calorimeter (PEM) covers the 
% region $1.1 < |\eta| < 3.6$, while the new hadronic calorimeter (PHA) provides 
% coverage in the $1.3 < |\eta| < 3.6$ region. 
%
The  
calorimeters are instrumented with  finely segmented detectors~\cite{ccal,pes} to measure 
the shower profile at a longitudinal depth close to the location of a typical electromagnetic 
shower maximum. 
Cherenkov counters in the region $3.7 < |\eta| < 4.7$ measure the number of 
inelastic $p \overline{p}$ collisions to compute the luminosity~\cite{clc}.

% ============================
%\section{Monte Carlo Simulation}   (DONE)
% ============================
Monte Carlo event samples are used to 
determine detector acceptance and reconstruction efficiency,
estimate background contributions, and unfold the measurements back to the hadron level~\cite{hadron}.
Samples of simulated inclusive $\Zee$+jets  events 
have been generated using the {\sc pythia} 6.216~\cite{pythia} Monte Carlo generator.
CTEQ5L~\cite{cteq5l} parton distribution functions (PDFs) are used for the proton and antiproton.
%
% The {\sc pythia} samples have been created using two special tuned sets of parameters, denoted as
% {\sc pythia-tune~a}~\cite{underlying} and {\sc pythia-tune~dw}~\cite{tunedw}, that include  
% enhanced contributions from initial-state gluon radiation and  secondary parton interactions 
% between proton and antiproton beam remnants and provide an accurate  description of the measured jet 
% shapes and energy flows in $\Zee$+jets final states~\cite{oriol}. 
%
The {\sc pythia} samples have been created using a special tuned set of parameters, denoted as
{\sc pythia-tune~a}~\cite{underlying}, that includes
enhanced contributions from initial-state gluon radiation and  secondary parton interactions
between proton and antiproton beam remnants and provides an accurate  description of the measured jet
shapes and energy flows in $\Zee$+jets final states~\cite{oriol}. 
% Fragmentation into hadrons is carried
% out using the string model~\cite{string}.
%
Monte Carlo samples for  
% $\tt$, $\Zee + \gamma$, dibosons and $\Ztt$+jets  
background processes are generated using {\sc pythia-tune~a}.
The samples are passed through a full
CDF detector simulation (based on  {\sc geant3}~\cite{geant} where the {\sc gflash}~\cite{gflash}
package is used to simulate the energy deposition in the calorimeters) and reconstructed 
and analyzed with the same analysis chain as
for the data.

% ============================
% Event selection   (DONE)
% ============================
% The measurements presented in this letter correspond to a total integrated luminosity of 
% 1.7$\pm$0.1~fb${}^{-1}$. 
Events are collected using a three-level trigger system~\cite{trigger}. 
At the first-level trigger, events are required to have a central electromagnetic calorimeter 
cluster ($|\eta| < 1 $) with $E_T$~\cite{coord}  above 8~GeV and an associated track  
with $ {p^{\rm track}_{\rm T}}$  above 8~GeV/$c$. Similarly, at the 
second-level (third-level) trigger a central electromagnetic cluster with $E_T>16$~GeV ($E_T>18$~GeV)
and an associated track with ${p^{\rm track}_{\rm T}}>8$~GeV/$c$ (${p^{\rm track}_{\rm T}}> 9$~GeV/$c$) 
are required. 
% At the third-level trigger, events are selected to have a central electromagnetic 
% cluster with $E_T>18$~GeV and an associated track with ${p^{\rm track}_{\rm T}}$ above 9~GeV/$c$.  
%
%
The events are then required to have two electrons~\cite{electron} with $E_T^e > 25$~GeV and a
reconstructed invariant mass in the range $66 < M_{ee} < 116$~GeV/$c^2$ around the $Z$ boson mass.  The 
electron candidates are reconstructed using criteria described in~\cite{ecuts}.
% with no isolation requirement applied to the electromagnetic clusters.
In this study, one electron is required to be central ($|\eta^e|<1$) and fulfill tight selection cuts,
while the second electron is required to pass a looser selection  and to be either central (CC final-state
configuration) or forward  with $1.2 < |\eta^e| < 2.8$ (CF final-state configuration).
The trigger efficiencies for CC and CF configurations are $99.96 \pm 0.01 \%$ and $97.9 \pm 0.3 \%$,
respectively. The events are selected to have a reconstructed primary vertex with $z$-position within
60~cm around the nominal interaction point, and at least one jet with corrected transverse momentum 
$\ptjetcor > 30$~GeV/$c$ (see below), rapidity~\cite{coord} in the range $|\yjetcal|<2.1$, and  $\Delta R_{e-jet} > 0.7$, where $\Delta R_{e-jet}$ denotes the distance  
($y - \phi$  space) between the jet and each of the two  electrons in the final state. The final sample contains 6203, 650, 57, and 2 events with
at least one, two, three, and four jets, respectively.

% ============================
% Jet reconstruction  & Correction  (DONE)
% ============================

Jets are reconstructed in data and Monte Carlo simulated events 
from the energy deposits in the calorimeter towers with transverse momenta~\cite{e-p}
above 0.1~GeV/$c$. The towers associated with the reconstructed electrons in the final state  
are excluded from the jet search. Jets are searched for using the midpoint algorithm~\cite{midpoint} with cone radius $R=0.7$ and a merging/splitting 
fraction of 0.75, starting from seed towers with transverse momenta above 1~GeV/$c$.
% The measured jet transverse momentum, rapidity, and azimuthal angles are denoted as $\ptjetcal$, $\yjetcal$, 
% and $\fjetcal$, respectively. 
The same algorithm is applied to the final state particles 
in the Monte Carlo generated events, excluding $Z/\gamma^*$ decay products, to define jets at the hadron level. 
% The resulting hadron-level jet variables are denoted as $\ptjethad$, $\yjethad$ and $\fjethad$.

The rapidity and azimuthal angle of the jets, $\yjetcal$ and $\fjetcal$, are well reconstructed in the calorimeter with 
a resolution better than 0.05 units in both $y$ and $\phi$. The 
measured jet transverse momentum $\ptjetcal$ systematically underestimates that of the hadron-level jet. For $\ptjetcal$ values about 30~GeV/$c$, 
the jet transverse momentum is underestimated by about 30$\%$. 
The systematic shift decreases with increasing $\ptjetcal$ down to about 11$\%$ for $\ptjetcal > 200$~GeV/$c$. This is mainly 
attributed to the presence of inactive material and the 
non-compensating nature of the calorimeter~\cite{calor}. An average correction, as a function of $\ptjetcal$ and $\yjetcal$, 
is applied to the measured 
$\ptjetcal$ to account for these effects~\cite{jetcornim}. The measured $\ptjetcal$ also includes contributions
from  multiple $p\bar{p}$ interactions per crossing at high instantaneous luminosity. 
Multiple interactions are identified via the presence of additional primary vertices reconstructed from charged particles.
For each jet, $\ptjetcal$ is
% The measured jet transverse momenta are 
corrected for  this effect by removing a certain amount of transverse momentum, $\deltapt = 1.06 \pm 0.32$~GeV/$c$, 
for each additional primary vertex in the event, as   determined from data~\cite{jetcornim}.
% A value $\deltapt  = 1.055 \pm 0.316$~GeV/$c$ is determined from data~\cite{jetcornim}.
%
% ============================
% Background substraction
% ============================

The main backgrounds to the $\Zee$+jets  sample arise from inclusive-jets and $W$+jets events, and 
are estimated from the data. First, an inclusive jet data sample  
is employed to estimate the probability $\fake$ for a jet to pass a given electron selection. 
The probabilities are parametrized as a function of $\ptjetcal$ and are 
typically around  
0.02  and 0.005  for central  and forward loose electrons, respectively. 
Second, a sample of events in data with exactly one reconstructed 
tight central electron is selected. For 
each jet in the event, the  $E_T^e$ of a fake electron is determined, and the 
invariant mass of the tight-central electron and jet final state is then computed. 
Event-by-event, all electron-jet combinations that fulfill the $E_T^e$ cuts and with an invariant mass
within $66 < M_{e-jet} < 116$~GeV/$c^2$ are considered in the background calculation,  
where each combination is weighted by the corresponding $\fake$ value and divided by the number 
of accepted electron-jet combinations in the event. The total inclusive-jets and $W$+jets background is 
then computed in each measured distribution.
Other background  contributions from $\tt$, $\Zee + \gamma$, $WW$, $WZ$, $ZZ$, 
and $\Ztt$+jets final states are estimated using Monte Carlo samples. 
The total background in  inclusive $\Zee$+jets production 
is about 12$\%$ for $\njet \geq 1$, and increases up to about 17$\%$ for $\njet \geq 3$.
Good agreement is observed in the total number of events between the data and 
the $\Zee$+jets signal plus background predictions.
% and in 
% $\pm 20$~GeV/$c^2$ sidebands around the selected $M_{ee}$ range. 
A $\chi^2$ test, where only statistical uncertainties are considered, gives  $\chi^2$ probabilities that vary between
$80 \%$ and $25 \%$ as $\njet$ increases.   

% ============================
% Unfolding          (DONE)
% ============================
% The measurements  are corrected back to the hadron level
% using {\sc pythia-tune~a}  event samples. The measurements refer to hadron level 
% jets with $\ptjet > 30$~GeV/$c$ and  $|\yjet|<2.1$, in 
% limited and well-defined kinematic range for the $Z$ decay products, as already discussed.
%  : 
% $E^e_T > 25$~GeV, $|\eta^{e1}| < 1.0$, $|\eta^{e2}|<1.0$ or
% $1.2 < |\eta^{e2}|<2.8$, $66 < M_{ee} < 116$~GeV/$c^2$, and  $\Delta R_{e-jet} > 0.7$.  
% the measurements  are corrected for acceptance and smearing effects using a bin-by-bin unfolding procedure, 
% which also accounts for the efficiency of the $\Zee$ selection criteria.  
% in order to avoid
% any bias on the correction factors due to
% the particular PDF set used,
% which translates into
% slightly different simulated $\ptjetcal$ distributions, the {\sc pythia-tune~a}  Monte Carlo event sample is 
% re-weighted until it accurately follows
% the measured $\ptjetcal$ spectra.
% the unfolding factors are computed separately 
% for the different inclusive $\Zee$+jets measurements and vary between 
% 2.0 at low~$\ptjet$ and 2.3 at high~$\ptjet$. 

% =================================
% UNFOLDING
% =================================

Raw inclusive jet differential cross sections as a function of $\ptjetcor$ are defined as 
$d\sigma/d\ptjetcor = \frac{1}{\mathcal{L}} \times (\njetcor/\Delta\ptjetcor)$, where 
$\njetcor$ denotes the total number of jets in a given $\ptjetcor$ bin, $\Delta\ptjetcor$ is the 
size of the bin, and $\mathcal{L}$ is the luminosity. $\njetcor$ is corrected bin-by-bin for background 
contributions and trigger inefficiencies. 
The measured cross sections are then corrected for acceptance and smearing effects  back to the hadron level  
using {\sc pythia-tune~a}  Monte Carlo event samples,  and a bin-by-bin 
unfolding procedure that also accounts for the efficiency of 
the $\Zee$ selection criteria. 
The final results refer to hadron level
jets with $\ptjet > 30$~GeV/$c$ and  $|\yjet|<2.1$, in 
a limited and well-defined kinematic range for the $Z/\gamma^*$ decay products:
%, as already discussed.
$E^e_T > 25$~GeV, $|\eta^{e1}| < 1.0$, $|\eta^{e2}|<1.0$ or
$1.2 < |\eta^{e2}|<2.8$, $66 < M_{ee} < 116$~GeV/$c^2$, and  $\Delta R_{e-jet} > 0.7$.
In order to avoid any bias on the correction factors due to the particular PDF set used,
which translates into slightly different simulated $\ptjetcal$ distributions, 
the {\sc pythia-tune~a}  Monte Carlo event sample is  re-weighted until it accurately follows
the measured $\ptjetcal$ spectra.
The unfolding factors
$U(\ptjetcor) = \frac{\rm d \sigma}{d \ptjet}/\frac{\rm d \sigma}{\rm d \ptjetcor}$ 
are computed separately for the different 
% inclusive $\Zee$+jets 
measurements and vary between 
2.0 at low~$\ptjet$ and 2.3 at high~$\ptjet$.

%
% ============================
% Syst. Uncertainties  (DONE)
% ============================
%

A detailed study of the systematic uncertainties was carried out~\cite{oriol}.
A $\pm 1.5 \%$ uncertainty  on the trigger efficiency translates into $\pm 1.5\%$  and $\pm 0.06\%$
uncertainties on the cross sections for CF and CC configurations, respectively. The uncertainty 
on the $\ptjet$ dependence of the electron identification efficiency introduces a $\pm 5 \%$ 
uncertainty on both CC and CF results.  
The measured jet  energies are varied by  $\pm 2 \%$ at low $\ptjetcal$ to $\pm 2.7 \%$ at high $\ptjetcal$
to account for the uncertainties on the absolute energy scale in
the calorimeter~\cite{jetcornim}; this introduces 
uncertainties on the final measurements which vary between 
$\pm 5 \%$ at low $\ptjet$  and $\pm 12 \%$
at high $\ptjet$. The $\yjet$ dependence of the average correction applied to $\ptjetcal$
introduces a $\pm 2\%$ uncertainty on the measured cross sections, approximately independent of $\ptjet$. 
The uncertainty on $\deltapt$ has a negligible effect on the measured cross sections. 
The uncertainty on the $\ptjetcal$ dependence of $\fake$ 
introduces a $\pm 15\%$ uncertainty on the inclusive-jets
and $W$+jets background estimation, that translates into a less than $2\%$ uncertainty on the 
measured cross sections.  A conservative  $\pm 30\%$ uncertainty on the normalization of the rest of the background 
contributions, as extracted from Monte Carlo samples, introduces a less than 1$\%$ effect on the 
final results. If the unfolding procedure is carried out using unweighted {\sc pythia-tune~a},  
the effect on the measured cross sections is less than 1$\%$. 
% The measurements are performed in different periods of Tevatron instantaneous luminosity 
% to account for possible 
% remaining contributions from pile-up events, and no significant effect is found. 
Positive and negative deviations with respect to the nominal cross section values are added 
separately in quadrature to define the total systematic uncertainty.  The final results are obtained from the combination 
of CC and CF measurements. 
Finally, a $5.8 \%$ uncertainty on the total 
luminosity is included in the measured cross sections.
% ============================
% RESULT 1
% ============================

\begin{figure}
\includegraphics[width=\figsize]{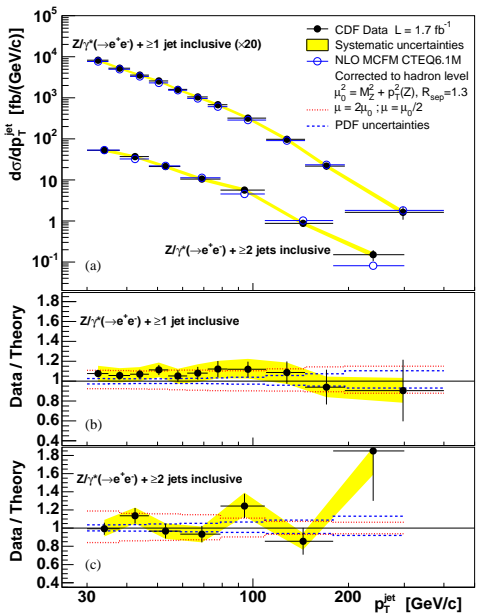}
\caption{(a) Measured inclusive jet differential cross section 
as a function of $\ptjet$
(black dots) in $\Zee$+jets  with 
$\njet \geq 1,2$ compared to NLO pQCD predictions (open circles). 
For clarity, the measurement for $\njet \geq 1$ is scaled up by ($\times$20).
The shaded bands show the total systematic uncertainty,
except for the 5.8$\%$ luminosity uncertainty.
(b and c) Data/theory ratio  as a function of $\ptjet$ for $\njet \geq 1$ and $\njet \geq 2$, respectively.
The dashed and dotted lines indicate the PDF uncertainty and the variation with $\mu$ of 
the NLO pQCD predictions, respectively.
}
\label{fig:pt}
\end{figure}
Figure~\ref{fig:pt}(a) shows the measured inclusive jet differential cross sections  as a function 
of $\ptjet$ in $\Zee$+jets production, with $\njet \geq 1$ and $\njet \geq 2$,   
compared to NLO pQCD predictions.  
The data are reported in Table~I.
The cross sections decrease by more than three orders of magnitude as $\ptjet$ increases 
from  30~GeV/c up to about 300~GeV/c. The NLO pQCD predictions are computed using the 
{\sc mcfm} program~\cite{mcfm} with CTEQ6.1M PDFs~\cite{cteq},  with the 
renormalization and factorization scales  set to $\mu^2  =  M_Z^2 + p^2_T(Z)$, and using a midpoint 
algorithm with $R=0.7$ and $R_{\rm sep} = 1.3$~\cite{rsep} to reconstruct jets at the parton level.
Values for $R_{\rm sep}$ between 1.0 and 2.0 change the theoretical prediction by less than $2\%$.
A variation of $\mu$ by a factor two (half) reduces (increases) the theoretical predictions by 
10$\%$ to 15$\%$. The uncertainties on the 
NLO pQCD predictions due to the PDFs were computed using the 
Hessian method~\cite{hessian}. They vary from $\pm 4\%$ at low $\ptjet$ to 
$\pm 10 \%$ at high $\ptjet$.  

The theoretical predictions include  parton-to-hadron  
correction  factors, $C_{\rm had}(\njet, \ptjet)$,  that approximately account for 
non-perturbative contributions from  the underlying event
and fragmentation into hadrons (see Table~I). 
In each measurement $C_{\rm had}$ is estimated using the {\sc pythia-tune~a} Monte Carlo samples, as the ratio between the 
nominal $\ptjet$ distribution and the one obtained 
by turning off both the interactions between proton and antiproton remnants and the 
string fragmentation in the Monte Carlo samples.  The correction decreases as $\ptjet$ increases from
about 1.2 (1.26) at $\ptjet$ of 30~GeV/c  to 
1.02 (1.01) for  $\ptjet > 200$~GeV/$c$ for $\njet \geq 1$ ($\njet \geq 2$), and is  
dominated by the underlying event contribution. 
In order to estimate the uncertainty on $C_{\rm had}$, {\sc pythia} samples
are generated with a different set of parameters, denoted as 
{\sc tune~dw}~\cite{tunedw}, that governs the underlying event activity 
and also describes the $\Zee$+jets final states. 
The uncertainty on $C_{\rm had}$ is about 
$10 \%$ ($17\%$) at low $\ptjet$ and goes down to $1 \%$ at high $\ptjet$
for $\njet \geq 1$ ($\njet \geq 2$).
%, as determined using the {\sc pythia-tune~dw} Monte Carlo sample
% instead of the {\sc pythia-tune~a} sample. 
%
% During the course of the analysis additional 
% $\Zee$=jets Monte Carlo samples generated with {\sc herwig}~6.5~\cite{herwig} interfaced with 
% {\sc JIMMY}~4.1~\cite{jimmy} were considered to determine $C_{\rm had}$.     
%
%
The ratios  between data and theory as a function of $\ptjet$ are shown in Fig.~\ref{fig:pt}(b,c). 
Good agreement is observed between the 
measured cross sections and the nominal theoretical predictions. A $\chi^2$ test, where the sources 
of systematic uncertainty on the data are considered independent but fully correlated across $\ptjet$ bins, and the 
uncertainty on $C_{\rm had}$ is also included, gives a $\chi^2$ probability of $99 \%$ ($22 \%$) for 
$\njet \geq 1$  ($\njet \geq 2$).

\begin{figure}[h]
\includegraphics[width=\figsize]{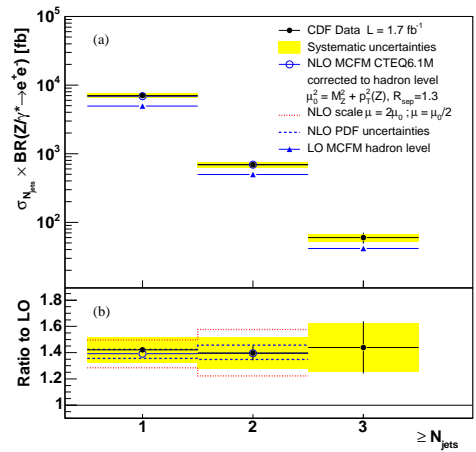}
\caption{ (a) Measured total cross section for inclusive jet production in $\Zee$ events as a function of $\njet$ compared
to LO and NLO pQCD predictions.  
The shaded bands show the total systematic uncertainty,
except for the 5.8$\%$ luminosity uncertainty.
(b)
Ratio of data and NLO to LO pQCD predictions versus  $\njet$.
The dashed and dotted lines indicate the PDF uncertainty and the variation with $\mu$ of 
the NLO pQCD predictions, respectively.
}
\label{fig:njet}
\end{figure}

% =======================================
% TABLE
% =======================================

\begin{table}[htbp] 
\begin{footnotesize} 
\begin{center} 
\renewcommand{\baselinestretch}{1.2} 
\begin{tabular}{|c|c|c|} \hline
$\ptjet$&  $\frac{\rm{d} \sigma}{\rm{d} \ptjet} \pm {\rm (stat.)} \pm {\rm (syst.)} \pm {\rm (lum.)}$ & $C_{\rm had}
\pm {\rm (stat.)} \pm {\rm (syst.)}$ \\
$[\rm GeV/c]$ & $[\rm fb/(GeV/c)]$ & parton $\rightarrow$ hadron\\ \hline\hline
\multicolumn{3}{|c|}{\small{$\Zee$+jets \ \ ($\njet \geq 1$)}} \\ \hline\hline
30 - 35 &  $413.3 \pm 13.3 {}^{+30.4}_{-31.3} \pm 24.0 $ & $1.209 \pm 0.010 \pm 0.134$\\ 
35 - 41 &  $263.3 \pm 9.4 {}^{+18.3}_{-17.4}  \pm 15.3 $ &  $1.146 \pm 0.010 \pm 0.096$\\
41 - 47 &  $178.3 \pm 7.5 {}^{+12.0}_{-11.6}  \pm 10.3 $ &  $1.114 \pm 0.011 \pm 0.077$\\ 
47 - 54 &  $128.5 \pm 5.9 {}^{+8.7}_{-8.4}    \pm  7.5 $ &    $1.097 \pm 0.012 \pm 0.066$\\ 
54 - 62 &  $ 80.5 \pm 4.3 {}^{+5.5}_{-6.0}    \pm  4.7 $ &    $1.086 \pm 0.013 \pm 0.059$\\ 
62 - 72 &  $ 52.5 \pm 3.2 {}^{+4.4}_{-4.3}    \pm  3.0 $ &    $1.078 \pm 0.013 \pm 0.053$\\ 
72 - 83 &  $ 34.2 \pm 2.4 {}^{+2.5}_{-2.8}    \pm  2.0 $ &    $1.072 \pm 0.015 \pm 0.049$\\ 
83 - 110 & $ 16.0 \pm 1.1 {}^{+1.5}_{-1.3}    \pm  0.9 $ &    $1.063 \pm 0.012 \pm 0.043$\\ 
110 - 146 &  $4.9 \pm 0.5 {}^{+0.5}_{-0.5}    \pm  0.3 $ &    $1.051 \pm 0.012 \pm 0.035$\\ 
146 - 195 &  $1.1 \pm 0.2 {}^{+0.1}_{-0.1}    \pm  0.06$ &    $1.040 \pm 0.008 \pm 0.027$\\ 
195 - 400 &  $0.08 \pm 0.03 {}^{+0.01}_{-0.01}\pm 0.005 $ &$1.021 \pm 0.005 \pm 0.013$\\ \hline\hline 
\multicolumn{3}{|c|}{\small{$\Zee$+jets \  \  ($\njet \geq 2$)}} \\ \hline\hline
30 - 38 & $52.9 \pm 3.5 {}^{+5.3}_{ -4.6} \pm 3.1 $ & $1.262 \pm 0.022 \pm 0.217$\\ 
38 - 47 & $37.0 \pm 2.8 {}^{+2.9}_{ -2.8} \pm 2.1 $ & $1.207 \pm 0.024 \pm 0.169$\\ 
47 - 59 & $21.2 \pm 1.8 {}^{+1.9}_{ -1.9} \pm 1.2 $ & $1.164 \pm 0.025 \pm 0.130$\\ 
59 - 79 & $10.5 \pm 1.0 {}^{+0.9}_{ -1.0} \pm 0.6 $ & $1.123 \pm 0.024 \pm 0.093$\\ 
79 - 109 & $5.7 \pm 0.6 {}^{+0.7}_{ -0.5} \pm 0.3 $ & $1.087 \pm 0.026 \pm 0.062$\\ 
109 - 179 & $0.88 \pm 0.15 {}^{+0.09}_{ -0.10} \pm  0.05  $ & $1.052 \pm 0.020 \pm 0.030$\\ 
179 - 300 & $0.15 \pm 0.04 {}^{+0.02}_{ -0.02} \pm  0.009 $ & $1.026 \pm 0.010 \pm 0.008$\\ \hline\hline
 \end{tabular} 
\end{center}  
\renewcommand{\baselinestretch}{1} 
\end{footnotesize}  
\label{tab:table}   
\caption{Measured inclusive jet differential cross section in $\Zee$+jets production 
as a function of $\ptjet$ with  $\njet \geq 1$ and $\njet \geq 2$. 
% The $5.8 \%$ uncertainty on the total
% luminosity is not included. 
The systematic uncertainties are fully correlated across $\ptjet$ bins. 
The parton-to-hadron
 correction
factors $C_{\rm had} (\ptjet,\njet)$ are applied to the pQCD predictions. 
}
\end{table}

Figure~\ref{fig:njet} shows the total cross sections, $\sigmanjet$,  for $\Zee$+jets  events
up to $\njet \geq  3$. The measured event cross sections are:  
{\small{$\sigma_{1} = 7003 \pm 146 \rm (stat.) {}^{+483}_{-470} (\rm syst.) \pm 406 (\rm lum.) \ \rm fb$}},
{\small{$\sigma_{2} =  695 \pm 37  \rm (stat.) {}^{ +59}_{ -60} (\rm syst.) \pm 40  (\rm lum.)  \ \rm fb$}}, and
{\small{$\sigma_{3} =   60 \pm 11  \rm (stat.) {}^{  +8}_{  -8} (\rm syst.) \pm 3.5 (\rm lum.)  \ \rm fb$}},
for $\njet \geq 1$, $\njet \geq 2$, and $\njet \geq  3$, respectively.
% The additional $5.8 \%$ uncertainty on the total
% luminosity is not included.
The data are compared to LO and NLO pQCD predictions. The  
parton-to-hadron non-perturbative corrections vary between 1.1 and 1.4 as $\njet$ increases.  
For $\njet \geq 1$ and $\njet \geq 2$, the LO pQCD predictions underestimate 
the  measured cross sections by a factor about 1.4, which corresponds to $\chi^2$ probabilities
of 0.07$\%$ and 2.7$\%$, respectively.
% smaller than $3 \%$. 
Good agreement is observed between data and NLO pQCD 
predictions, with $\chi^2$ probabilities better than $83 \%$. For $\njet \geq 3$, where no NLO pQCD prediction is  available, 
the measured cross section indicates that the data 
can be described by  a LO-to-NLO theoretical  factor~independent~of~$\njet$. 

% =======================================
% TABLE INTEGRATED
% =======================================

% \begin{table}[htbp]
% \begin{footnotesize}
% \begin{center}
% \renewcommand{\baselinestretch}{1.2}
% \begin{tabular}{|c|c|c|} \hline
% $\njet$ & $\sigma \pm {\rm (stat.)} \pm {\rm (sys.)} \ [\rm fb]$ & $C_{\rm had}  \pm {\rm (stat.)} \pm {\rm (sys.)}$ \\ \hline
%  1 & 6981   $\pm 153 {}^{+481}_{-468}$ & 1.137 $\pm$  0.005 $\pm$ 0.072 \\  
%  2 & 684    $\pm 40  {}^{+59}_{-59}$   &  1.192 $\pm$  0.016 $\pm$ 0.109 \\ 
%  3 &  59    $\pm 11  {}^{+8}_{-8}$     &  1.367 $\pm$  0.069 $\pm$ 0.325 \\ \hline
% \end{tabular}
% \end{center}
% \renewcommand{\baselinestretch}{1}
% \end{footnotesize}
% \label{tab:tableii}
% \caption{Measured total cross section for $\Zeejets$ production as a function of $\njet$. 
% A $5.8 \%$ uncertainty on the total
% luminosity is not included.
% The parton-to-hadron correction factors, $C_{\rm had} (\njet)$, are applied to the pQCD predictions.}
% \end{table}

% ========================
% SUMMARY AND CONCLUSIONS
% ========================

%\section{Conclusion}
In summary, we report new results on inclusive jet production in $\Zee$ events 
in $p \overline{p}$ collisions at $\sqrt{s} = 1.96 \ {\rm TeV}$ 
for jets with    
$\ptjet > 30$~GeV/c and $| \yjet |<2.1$,  
based on $1.7 \ \rm fb^{-1}$ of CDF Run~II data.
The measured cross sections are well described by NLO pQCD predictions including non-perturbative~corrections. 

%\begin{center}
%\textbf{Acknowledgments}
%\end{center}

We thank the Fermilab staff and the technical staffs of the participating institutions for 
their vital contributions. This work was supported by the U.S. Department of Energy and National 
Science Foundation; the Italian Istituto Nazionale di Fisica Nucleare; the Ministry of 
Education, Culture, Sports, Science and Technology of Japan; the Natural Sciences and 
Engineering Research Council of Canada; the National Science Council of the Republic of China; the Swiss 
National Science Foundation; the A.P. Sloan Foundation; the Bundesministerium f\"ur Bildung und 
Forschung, Germany; the Korean Science and Engineering Foundation and the Korean Research 
Foundation; the Science and Technology Facilities Council and the Royal Society, UK; the 
Institut National de Physique Nucleaire et Physique des Particules/CNRS; the Russian Foundation 
for Basic Research; the Comisi\'on Interministerial de Ciencia y Tecnolog\'{\i}a, Spain; 
the European Community's Human Potential Programme; the Slovak R$\&$D Agency; and the Academy of Finland.

\end{document}

%% file: July_2007_Authors_Visitors.tex
\affiliation{Institute of Physics, Academia Sinica, Taipei, Taiwan 11529, Republic of China} 
\affiliation{Argonne National Laboratory, Argonne, Illinois 60439} 
\affiliation{Institut de Fisica d'Altes Energies, Universitat Autonoma de Barcelona, E-08193, Bellaterra (Barcelona), Spain} 
\affiliation{Baylor University, Waco, Texas  76798} 
\affiliation{Istituto Nazionale di Fisica Nucleare, University of Bologna, I-40127 Bologna, Italy} 
\affiliation{Brandeis University, Waltham, Massachusetts 02254} 
\affiliation{University of California, Davis, Davis, California  95616} 
\affiliation{University of California, Los Angeles, Los Angeles, California  90024} 
\affiliation{University of California, San Diego, La Jolla, California  92093} 
\affiliation{University of California, Santa Barbara, Santa Barbara, California 93106} 
\affiliation{Instituto de Fisica de Cantabria, CSIC-University of Cantabria, 39005 Santander, Spain} 
\affiliation{Carnegie Mellon University, Pittsburgh, PA  15213} 
\affiliation{Enrico Fermi Institute, University of Chicago, Chicago, Illinois 60637} 
\affiliation{Comenius University, 842 48 Bratislava, Slovakia; Institute of Experimental Physics, 040 01 Kosice, Slovakia} 
\affiliation{Joint Institute for Nuclear Research, RU-141980 Dubna, Russia} 
\affiliation{Duke University, Durham, North Carolina  27708} 
\affiliation{Fermi National Accelerator Laboratory, Batavia, Illinois 60510} 
\affiliation{University of Florida, Gainesville, Florida  32611} 
\affiliation{Laboratori Nazionali di Frascati, Istituto Nazionale di Fisica Nucleare, I-00044 Frascati, Italy} 
\affiliation{University of Geneva, CH-1211 Geneva 4, Switzerland} 
\affiliation{Glasgow University, Glasgow G12 8QQ, United Kingdom} 
\affiliation{Harvard University, Cambridge, Massachusetts 02138} 
\affiliation{Division of High Energy Physics, Department of Physics, University of Helsinki and Helsinki Institute of Physics, FIN-00014, Helsinki, Finland} 
\affiliation{University of Illinois, Urbana, Illinois 61801} 
\affiliation{The Johns Hopkins University, Baltimore, Maryland 21218} 
\affiliation{Institut f\"{u}r Experimentelle Kernphysik, Universit\"{a}t Karlsruhe, 76128 Karlsruhe, Germany} 
\affiliation{Center for High Energy Physics: Kyungpook National University, Daegu 702-701, Korea; Seoul National University, Seoul 151-742, Korea; Sungkyunkwan University, Suwon 440-746, Korea; Korea Institute of Science and Technology Information, Daejeon, 305-806, Korea; Chonnam National University, Gwangju, 500-757, Korea} 
\affiliation{Ernest Orlando Lawrence Berkeley National Laboratory, Berkeley, California 94720} 
\affiliation{University of Liverpool, Liverpool L69 7ZE, United Kingdom} 
\affiliation{University College London, London WC1E 6BT, United Kingdom} 
\affiliation{Centro de Investigaciones Energeticas Medioambientales y Tecnologicas, E-28040 Madrid, Spain} 
\affiliation{Massachusetts Institute of Technology, Cambridge, Massachusetts  02139} 
\affiliation{Institute of Particle Physics: McGill University, Montr\'{e}al, Canada H3A~2T8; and University of Toronto, Toronto, Canada M5S~1A7} 
\affiliation{University of Michigan, Ann Arbor, Michigan 48109} 
\affiliation{Michigan State University, East Lansing, Michigan  48824} 
\affiliation{University of New Mexico, Albuquerque, New Mexico 87131} 
\affiliation{Northwestern University, Evanston, Illinois  60208} 
\affiliation{The Ohio State University, Columbus, Ohio  43210} 
\affiliation{Okayama University, Okayama 700-8530, Japan} 
\affiliation{Osaka City University, Osaka 588, Japan} 
\affiliation{University of Oxford, Oxford OX1 3RH, United Kingdom} 
\affiliation{University of Padova, Istituto Nazionale di Fisica Nucleare, Sezione di Padova-Trento, I-35131 Padova, Italy} 
\affiliation{LPNHE, Universite Pierre et Marie Curie/IN2P3-CNRS, UMR7585, Paris, F-75252 France} 
\affiliation{University of Pennsylvania, Philadelphia, Pennsylvania 19104} 
\affiliation{Istituto Nazionale di Fisica Nucleare Pisa, Universities of Pisa, Siena and Scuola Normale Superiore, I-56127 Pisa, Italy} 
\affiliation{University of Pittsburgh, Pittsburgh, Pennsylvania 15260} 
\affiliation{Purdue University, West Lafayette, Indiana 47907} 
\affiliation{University of Rochester, Rochester, New York 14627} 
\affiliation{The Rockefeller University, New York, New York 10021} 
\affiliation{Istituto Nazionale di Fisica Nucleare, Sezione di Roma 1, University of Rome ``La Sapienza," I-00185 Roma, Italy} 
\affiliation{Rutgers University, Piscataway, New Jersey 08855} 
\affiliation{Texas A\&M University, College Station, Texas 77843} 
\affiliation{Istituto Nazionale di Fisica Nucleare, University of Trieste/\ Udine, Italy} 
\affiliation{University of Tsukuba, Tsukuba, Ibaraki 305, Japan} 
\affiliation{Tufts University, Medford, Massachusetts 02155} 
\affiliation{Waseda University, Tokyo 169, Japan} 
\affiliation{Wayne State University, Detroit, Michigan  48201} 
\affiliation{University of Wisconsin, Madison, Wisconsin 53706} 
\affiliation{Yale University, New Haven, Connecticut 06520} 
\author{T.~Aaltonen}
\affiliation{Division of High Energy Physics, Department of Physics, University of Helsinki and Helsinki Institute of Physics, FIN-00014, Helsinki, Finland}
\author{J.~Adelman}
\affiliation{Enrico Fermi Institute, University of Chicago, Chicago, Illinois 60637}
\author{T.~Akimoto}
\affiliation{University of Tsukuba, Tsukuba, Ibaraki 305, Japan}
\author{M.G.~Albrow}
\affiliation{Fermi National Accelerator Laboratory, Batavia, Illinois 60510}
\author{B.~\'{A}lvarez~Gonz\'{a}lez}
\affiliation{Instituto de Fisica de Cantabria, CSIC-University of Cantabria, 39005 Santander, Spain}
\author{S.~Amerio}
\affiliation{University of Padova, Istituto Nazionale di Fisica Nucleare, Sezione di Padova-Trento, I-35131 Padova, Italy}
\author{D.~Amidei}
\affiliation{University of Michigan, Ann Arbor, Michigan 48109}
\author{A.~Anastassov}
\affiliation{Rutgers University, Piscataway, New Jersey 08855}
\author{A.~Annovi}
\affiliation{Laboratori Nazionali di Frascati, Istituto Nazionale di Fisica Nucleare, I-00044 Frascati, Italy}
\author{J.~Antos}
\affiliation{Comenius University, 842 48 Bratislava, Slovakia; Institute of Experimental Physics, 040 01 Kosice, Slovakia}
% ...
\collaboration{CDF Collaboration\footnote{With visitors from $^a$University of Athens, 15784 Athens, Greece, 
$^b$Chinese Academy of Sciences, Beijing 100864, China, 
$^c$University of Bristol, Bristol BS8 1TL, United Kingdom, 
$^d$University Libre de Bruxelles, B-1050 Brussels, Belgium, 
$^e$University of California Irvine, Irvine, CA  92697, 
$^f$University of California Santa Cruz, Santa Cruz, CA  95064, 
$^g$Cornell University, Ithaca, NY  14853, 
$^h$University of Cyprus, Nicosia CY-1678, Cyprus, 
$^i$University College Dublin, Dublin 4, Ireland, 
$^j$University of Edinburgh, Edinburgh EH9 3JZ, United Kingdom, 
$^k$University of Heidelberg, D-69120 Heidelberg, Germany, 
$^l$Universidad Iberoamericana, Mexico D.F., Mexico, 
$^m$University of Manchester, Manchester M13 9PL, England, 
$^n$Nagasaki Institute of Applied Science, Nagasaki, Japan, 
$^o$University de Oviedo, E-33007 Oviedo, Spain, 
$^p$Queen Mary, University of London, London, E1 4NS, England, 
$^q$Texas Tech University, Lubbock, TX  79409, 
$^r$IFIC(CSIC-Universitat de Valencia), 46071 Valencia, Spain, 
}}
\noaffiliation

%% file: z_final.bbl
\begin{thebibliography}{99}
\bibitem{coord} We use a cylindrical coordinate system about the beam axis 
with polar angle $\theta$ and azimuthal angle $\phi$.
We define $E_T = E \ {\rm sin} \theta$,  
$p_T = p \ {\rm sin}  \theta$, pseudorapidity $\eta = - {\rm ln(tan(\frac{\theta}{2})) }$, and rapidity  
$y = \rm \frac{1}{2} ln(\frac{E+p_z}{E-p_z}) $.
\bibitem{hadron}  The hadron level in the Monte Carlo generators is defined
using all final-state particles with lifetimes above $10^{-11} \ {\rm s}$.
\bibitem{electron} Charge conjugation is implied throughout the paper.
\bibitem{e-p} The momentum is computed using the energy and the position
with respect to the primary interaction vertex.
\end{thebibliography}
